\documentclass[aps,prb,twocolumn]{revtex4-1}
\usepackage[T1]{fontenc}
\usepackage[latin1]{inputenc}
\usepackage{lmodern}
\expandafter\let\csname equation*\endcsname\relax
\expandafter\let\csname endequation*\endcsname\relax
\usepackage{amsmath}
\usepackage{esint}
\usepackage{amssymb}
\usepackage{graphicx}
\usepackage{xcolor}
\usepackage{natbib}
\usepackage{bm}
\usepackage{bbold}
\usepackage[mathscr]{euscript}
\usepackage[cal=boondoxo]{mathalfa}
\usepackage{comment}

\def\ln{\textrm{ln}}

\def\pM{\mathrel{\raise 2pt \hbox{\tiny(}\!\raise 1pt \hbox{+}\settowidth {\dimen03} {+}\hskip-\dimen03 \raise -2.4pt \hbox {$-$} \!\raise 2pt \hbox{\tiny)}}}

\begin{document}
\title{Extended linear-in-$T$ resistivity due to electron-phason scattering in moir\'e superlattices}
\author{H\'ector Ochoa$^{1,2,3}$ and Rafael M. Fernandes$^4$}
\affiliation{$^1$Donostia International Physics Center, 20018 Donostia-San Sebastian, Spain\\
$^2$IKERBASQUE, Basque Foundation for Science, Maria Diaz de Haro 3, 48013 Bilbao, Spain\\
$^3$Department of Physics, Columbia University, New York, NY 10027, USA\\
$^4$School of Physics and Astronomy, University of Minnesota, Minneapolis, MN 55455, USA}

\date{\today}
\begin{abstract}
Due to its incommensurate nature, moir\'e superlattices host not only acoustic phonons but also another type of soft collective modes called phasons. Here, we investigate the impact of electron-phason scattering on the transport properties of moir\'e systems. We show that the resistivity can scale linearly with temperature down to temperatures much lower than the Bloch-Gr\"uneisen scale defined by electron kinematics on the Fermi surface. This result stems from the friction between layers, which transfers phason spectral weight to a broad diffusive low-energy peak in the mechanical response of the system. As a result, phason scattering becomes a very efficient channel for entropy production at low temperatures. We also consider the contributions of phasons to thermodynamic properties at low temperatures and find a ``metallic-like'' linear-in-$T$ behavior for the specific heat, despite the fact that this behavior is due to mechanical and not electronic degrees of freedom. We discuss the implications of this finding to reports of linear-in-$T$ resistivity in the phase diagram of twisted bilayer graphene.\end{abstract}
\maketitle

\section{Introduction}

Elucidating the nature of the metallic state of twisted moir\'e systems, from which correlated insulating and superconducting phases emerge,\cite{Cao2018a,Cao2018b,Yankowitz1059,Sharpe19,STM_Andrei19,STM_Pasupathy19,Efetov19,STM_Yazdani19,STM_Yazdani19,STM_Perge19,Young19,Pablo_nematics,Pomeranchuk1,Pomeranchuk2} is crucial to shed light on the microscopic ingredients governing the interplay between these phases.\cite{Xu_Balents2018,Yuan2018,Po2018,Dodaro2018,Kang2018,Guinea2018,Rademaker2018,Scalettar2018,Lin2018,PLee2018,Venderbos18,Sherkunov2018,Ochi2018,Kennes2018,Isobe2018,Bernevig_phonons,Tarnopolsky2019,Kang2019,Seo2019,Song2019,Bascones19,Pixley2019,Bernevig_TBGVI,Cenke2020,Cea2020,Christos20,Xie2020,Bultinck2020,Vafek_Kang_PRL_2020,Kim2021,Fernandes_ZYMeng,Song2021,Chichinadze2021,Khalaf2021,Kontani2022,Hofmann2022,Vafek_Wang2022} In the metallic phase of twisted bilayer graphene (TBG), puzzling features are seen both in its electronic spectrum, manifested as so-called cascade transitions,\cite{cascade,Wong2020} and in its transport properties. Indeed, while not exceeding $h/e^2$, relatively large resistivity values are observed, of about several k$\Omega$.\cite{transport1,transport2,transport3} Most strikingly, a resistivity that changes linearly with temperature is observed down to very low temperatures and over a wide range of carrier concentrations -- even when correlations are suppressed by screening.\cite{transport3}

On the one hand, this observation of a linear-in-$T$ resistivity is reminiscent of the phenomenology of strange metals, which are often associated with quantum critical points (QCP) in correlated electron systems.\cite{Taillefer2010,Bruin2013,Shibauchi2014} On the other hand, electron-acoustic phonon scattering is known to promote linear-in-$T$ resistivity down to the Bloch-Gr\"uneisen temperature, $T_{\textrm{BG}}$, or the Debye temperature, $T_{\textrm{D}}$.\cite{DasSarma,Adam1,Adam2,DasSarma2022} Both scenarios face difficulties: the fact that the linear-in-$T$ behavior extends over a broad doping range, rather than inside a cone emanating from a single point, is inconsistent with the standard QCP scenario. In the phonon scenario, the large in-plane rigidity and low mass density of the graphene layers leads to sound velocities $c_s\sim 10^4$ m/s, rendering the temperature scales $T_{\textrm{BG}}$ and $T_{\textrm{D}}$ relatively large compared to the temperatures for which linear-in-$T$ behavior is observed.


\begin{figure}[t!]
\begin{center}
\includegraphics[width=\columnwidth]{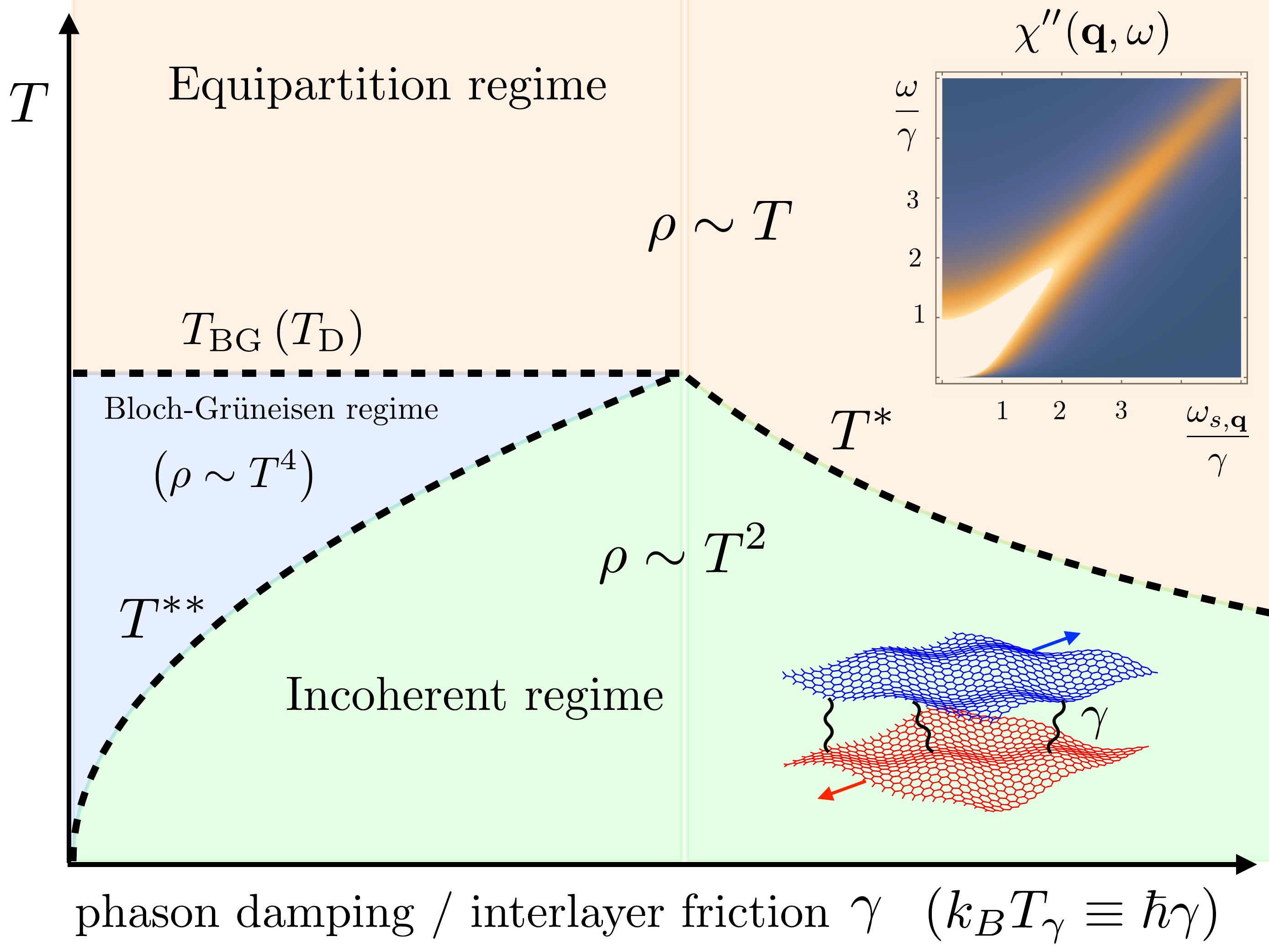}
\caption{Schematics of the temperature dependence of the resistivity due to electron-phason scattering. The horizontal axis is the phenomenological parameter $\gamma$ characterizing frictional forces between the layers, as schematically depicted in the lower inset. When those are absent, there is a single crossover from the high-temperature (i.e. classical equipartition) regime with linear-in-$T$ resistivity to the so-called Bloch-G\"uneisen regime, where $\rho \sim T^4$ (for a circular Fermi surface). For finite damping, low-energy phasons are overdamped and $\rho\sim T^2$ emerges at the lowest temperatures below $T^{**}$. As damping grows this scale saturates to $T_{\textrm{BG}}$ and the Bloch-G\"uneisen regime disappears, signaling that all scattering phason modes are overdamped. In this regime, there is a single crossover from linear- to quadratic-in-$T$ resistivity at $T^{*}<T_{\textrm{BG}}$. The upper inset represents the imaginary part of the phason susceptibility in Eq.~\eqref{eq:susceptibility}, characterized by a broad diffusive (i.e. incoherent) peak at low frequencies.} 
\label{fig:fig1}
\end{center}
\end{figure}

One important aspect of this problem that has remained little explored is the fact that, besides acoustic phonons emerging from the displacement of the center-of-mass of the bilayer, TBG and other moir\'e superlattices also possess another family of acoustic modes arising from the relative displacement between the layers. The latter describe the vibrations of the moir\'e pattern as a whole, and thus are sometimes dubbed \textit{moir\'e phonons}.\cite{Koshino} 
However, in contrast to conventional acoustic phonons, these modes are generally overdamped at long wavelengths since the relative momentum between the layers is not a conserved quantity. This is analogous to the phason excitations of incommensurate lattices.\cite{Winger,Rice,Rice_bis,Currat_etal} 
As such, because a moir\'e superlattice is generally an incommensurate lattice, these moir\'e modes have been identified as \textit{phasons}.\cite{phasonsI,phasons_TMD,phasons_Eslam,phasons_Matthias} Importantly, the dynamical mechanical response function $\chi_s$ of the bilayer at low frequencies is dominated by the two acoustic phason branches (transverse and longitudinal, labelled by $s$)  with dispersion $\omega_{s,\mathbf{q}}=c_{s}|\mathbf{q}|$ \cite{Koshino,phasonsI,phasons_TMD,phasons_Eslam,phasons_Matthias} and of the general form \cite{phasonsII}
\begin{align}
\label{eq:susceptibility}
\chi_{s}\left(\mathbf{q},\omega\right)=\frac{\varrho^{-1}}{\omega_{s,\mathbf{q}}^2-\omega^2-i\gamma\omega}\, .
\end{align}
Here, $\varrho$ (with units of mass density) is the inertia of the relative motion between the two layers, and $\gamma$ describes the damping of this motion due to frictional forces between the layers. 
While phasons have been widely studied in incommensurate lattices and quasicrystals,\cite{Chaikin_Lubensky} the impact of electron-phason scattering on the electronic properties of those systems has been relatively unexplored. Moir\'e superlattices, being correlated electronic systems, provide a unique framework to investigate this effect. 

In this article, we show that electron scattering by long-wavelength phason modes described by Eq.~\eqref{eq:susceptibility} can give rise to a linear-in-$T$ resistivity down to a new low-temperature scale $T^{*}\ll T_{\textrm{BG}}, T_{\textrm{D}}$. Figure~\ref{fig:fig1} summarizes our results for the different regimes for the phason-induced resistivity, obtained from Boltzmann-equation calculations. In the absence of interlayer friction, the resistivity $\rho$ displays the usual temperature dependence $\rho \sim T$ above $T_{\textrm{BG}}$ (or $T_{\textrm{D}}$) and $\rho \sim T^4$ below $T_{\textrm{BG}}$ (for a circular Fermi surface).\cite{DasSarma} For small damping, however, a second temperature scale $T^{**}$ emerges, below which the temperature dependence changes to $\rho \sim T^2$. This is a consequence of electrons scattering off of the phason modes associated with the low-energy diffusive peak of the response function (see inset in Fig.~\ref{fig:fig1}), and is reminiscent of the widely-studied case of scattering by overdamped bosonic fluctuations above a QCP.\cite{Moriya,Pines,RiceII,Rosch} Indeed, the phason propagator in Eq.~\eqref{eq:susceptibility} is similar to the bosonic propagator near a metallic QCP.\cite{Millis1993,Abanov2003,Sachdev2010,Fradkin2001,Lohneysen_etal}

When damping is further increased, $T^{**}$ overcomes $T_{\textrm{BG}}$, and essentially all relevant scattering phason modes are overdamped. In this situation, the $\rho \sim T^4$ behavior is completely suppressed, and the linear-in-$T$ behavior extends down to the new temperature scale $T^*$. Because scattering is no longer limited by the rigidity of individual graphene layers, but rather by the rate $\gamma$ at which the two layers exchange energy and momentum, this new temperature scale can be very small, $T^* \ll T_{\textrm{BG}}$ (see Eq.~\ref{eq:crossover}). Therefore, electron-phason scattering makes it possible for an extended regime of linear-in-$T$ resistivity in twisted moir\'e systems.

Based on this model, we expect the linear-in-$T$ resistivity of TBG to be accompanied by anomalous behaviors in other transport and thermodynamic properties at low temperatures due to the presence of low-energy phason excitations. This expectation is based on the similarity with the mechanical response of amorphous solids and glasses characterized by an excess of vibrational modes at low frequencies rooted in structural disorder and anharmonicity,\cite{Alexander1998,BP} which are also intrinsic to moir\'e systems.\cite{phasonsII} To illustrate this effect, we also compute the phason contribution to the specific heat at constant (hetero-)stress. We show that the specific heat is linear-in-$T$, $C_{\sigma}\propto T$, at low temperatures, $T\ll\textrm{min}\{ T_{\gamma},T_{\textrm{D}}^2/T_{\gamma} \}$, which dominates over the contribution from standard acoustic phonons ($\propto T^2$). This behavior is characteristic of intrinsically disordered systems, including incommensurate lattices, \cite{Bulaevskii1993,Cano_etal,Jiang2023} and should also impact other thermodynamic quantities.

The structure of the manuscript is as follows. We start in Sec.~\ref{sec:transport} from a general expression of the phason-limited resistivity within Boltzmann transport theory. Based on a relaxation-time approximation we discuss the different regimes in transport expected from the temperature dependence of the scattering rate. This expectation is confirmed in Sec.~\ref{sec:resistivity} by explicitly solving the Boltzmann equation. We provide analytic expressions for the Dirac approximation of the flat bands and numerical evaluations beyond the relaxation-time approximation in a tight-binding model. The phason contribution to the specific heat is evaluated in Sec.~\ref{sec:heat_capacity}. We conclude by summarizing our findings in Sec.~\ref{sec:conclusions}.

\section{Phason-limited electronic transport}

\label{sec:transport}

\subsection{Boltzmann transport theory}

In this work, we compute the resistivity within a Boltzmann transport approach. In the case of metallic TBG, this approach is justified by the empirical observation that the Mott-Ioffe-Regel limit is satisfied, i.e. the resistivity saturates when the mean-free-path becomes comparable to the Fermi wavelength, $k_F\ell\gtrsim 1$.\cite{transport3} 
The resistivity can be written as\cite{Ziman} \begin{align}
\label{eq:resistivity1}
\rho=\frac{1}{4e^2}\frac{\frac{1}{2k_BT}\int\frac{d\mathbf{k}_1}{(2\pi)^2}\int\frac{d\mathbf{k}_2}{(2\pi)^2}\mathcal{P}_{\mathbf{k_1},\mathbf{k}_2}\left(\Phi_{\mathbf{k}_1}-\Phi_{\mathbf{k}_2}\right)^2}{\left|\int\frac{d\mathbf{k}}{(2\pi)^2}\Phi_{\mathbf{k}}\,\boldsymbol{v}_{\mathbf{k}}\frac{\partial\, n_F}{\partial\varepsilon_{\mathbf{k}}}\right|^2},
\end{align}
where the factor of $4$ in the denominator arises from spin and valley degeneracies and $\Phi_{\mathbf{k}}$ solves the linearized Boltzmann equation in the presence of an electric field $\mathbf{E}$, \begin{align}
\label{eq:Boltzmann_eq}
-e\,\boldsymbol{v}_{\mathbf{k}}\cdot\mathbf{E}\,\frac{\partial\, n_F}{\partial\varepsilon_{\mathbf{k}}}=\frac{1}{k_B T}\int\frac{d\mathbf{k}'}{(2\pi)^2}\mathcal{P}_{\mathbf{k},\mathbf{k}'}\left(\Phi_{\mathbf{k}}-\Phi_{\mathbf{k}'}\right).
\end{align}

In these expressions, $\boldsymbol{v}_{\mathbf{k}}$ is the electron group velocity and $\mathcal{P}_{\mathbf{k_1},\mathbf{k}_2}$ represents the transition rate between states with momenta $\mathbf{k}_1$ and $\mathbf{k}_2$. Assuming that the lattice degrees of freedom relax much faster than the electron ensemble, and using detailed balance, the contribution to $\mathcal{P}_{\mathbf{k_1},\mathbf{k}_2}$ coming from electron-phason scattering processes can be written as\begin{align} \label{eq_P}
\mathcal{P}_{\mathbf{k_1},\mathbf{k}_2} & =2\left|g_s\left(\mathbf{k}_1,\mathbf{k}_2\right)\right|^2n_F\left(\varepsilon_{\mathbf{k}_1}\right)\left[1-n_F\left(\varepsilon_{\mathbf{k}_2}\right)\right]
\\
& \times \int_{-\infty}^{\infty}d\omega\, n_B(\omega)\chi_s''\left(\mathbf{k}_2-\mathbf{k}_1,\omega\right)\delta\left(\varepsilon_{\mathbf{k}_2}-\varepsilon_{\mathbf{k}_1}-\hbar\omega\right).
\nonumber
\end{align}
In this expression, $n_F$ and $n_B$ are Fermi-Dirac and Bose-Einstein distribution functions, respectively, $\chi_s''$ is the imaginary part of the susceptibility in Eq.~\eqref{eq:susceptibility}, and $g_s(\mathbf{k}_1,\mathbf{k}_2)$ represents the matrix element of the electron-phason coupling.

To simplify the analysis in this section, we consider a relaxation-time approximation, $\Phi_{\mathbf{k}}\propto\hat{\mathbf{u}}_{\mathbf{E}}\cdot\mathbf{k}$, where $\hat{\mathbf{u}}_{\mathbf{E}}$ is a unit vector along the external field. Later in Sec.~\ref{sec:resistivity} we will consider variational solutions of the Boltzmann equation beyond this approximation; the conclusions of the present analysis hold also in that case. At low temperatures, as long as the Fermi velocity is larger than the sound velocity,\cite{DasSarma2022} we expect that only electrons near the Fermi surface contribute to transport. Assuming that the resistivity is dominated by intraband processes, the resistivity can be approximated by\begin{align}
\label{eq:resistivity2}
\rho\approx \frac{\hbar}{2 e^2}\frac{\oint \frac{d\mathbf{k}_{\parallel}}{|\boldsymbol{v}_{\mathbf{k}}|}|\mathbf{k}|^2\tau_{\mathbf{k}}^{-1}}{\left[\oint \frac{d\mathbf{k}_{\parallel}}{|\boldsymbol{v}_{\mathbf{k}}|}\mathbf{k}\cdot\boldsymbol{v}_{\mathbf{k}}\right]^2},
\end{align}
where the integral is along the Fermi contour and the inverse of the transport time is given by 
\begin{align}
\tau_{\mathbf{k}}^{-1}=\oint \frac{d\mathbf{k}_{\parallel}'}{|\boldsymbol{v}_{\mathbf{k}'}|}\frac{\left|g_s\left(\mathbf{k},\mathbf{k}'\right)\right|^2}{\varrho k_B T}\frac{\left|\mathbf{k}-\mathbf{k}'\right|^2}{\left|\mathbf{k}\right|^2}f\left(\frac{\hbar \omega_{s,\mathbf{k}-\mathbf{k}'}}{k_BT},\frac{\hbar\gamma}{k_BT}\right).
\end{align}
The function $f\left(y,z\right)$ can be directly computed from the transition rate in Eq. \eqref{eq_P} and with $\chi_s$ from Eq. \eqref{eq:susceptibility}. We find
\begin{align}
\label{eq:f}
& f\left(y,z\right)=\frac{\pi}{y^2}+\frac{1}{4\pi\sqrt{z^2-4y^2}}\times\\\nonumber
&\left[\left(z-\sqrt{z^2-4y^2}\right)\psi_1\left(1+\frac{z-\sqrt{z^2-4y^2}}{4\pi}\right)\right.\\
&\left.-\left(z+\sqrt{z^2-4y^2}\right)\psi_1\left(1+\frac{z+\sqrt{z^2-4y^2}}{4\pi}\right)\right],\nonumber
\end{align}
where $\psi_1(x)$ is the trigamma function. 

There are \textit{a priori} two temperature scales in the problem associated with the two arguments of the function $f\left(y,z\right)$, which ultimately are connected to the poles of the susceptibility in Eq. \eqref{eq:susceptibility}. The first scale is determined by the maximum transferred momentum $\mathbf{k}-\mathbf{k}'$, which is limited either by the lattice (defining the Debye temperature $T_{\textrm{D}}$) or, for a small Fermi surface, as in doped TBG, by some multiple of the characteristic Fermi wavevector $k_F$. This is the Bloch-Gr\"uneisen temperature which, for a circular Fermi surface, is given by $k_BT_{\textrm{BG}}=2\hbar c_sk_F$. This scale is associated with underdamped phason oscillations, which take place above a characteristic momentum and
correspond to the sharp (i.e. coherent) part of the phason spectral weight shown in the inset of Fig. \ref{fig:fig1}.

However, for small momenta, the phason oscillations are overdamped, as shown by the low-energy incoherent phason spectral weight in the inset of Fig. \ref{fig:fig1}. They give rise to a second temperature scale, $k_BT_{\gamma}\equiv \hbar\gamma$, proportional to the rate of dissipation of energy and of relative linear momentum between the two layers. The relative strength of these two temperature scales define two distinct regimes of phason-limited transport: the \textit{propagating regime}, $T_{\textrm{BG}}\gg T_{\gamma}$, in which most of the phason modes scattering electrons behave as propagating waves, and the \textit{diffusive regime}, $T_{\textrm{BG}}\ll T_{\gamma}$, where most scattering modes are overdamped.

\subsection{Crossover temperature to linear-in-$T$ resistivity: Qualitative analysis}

Before computing the resistivity explicitly, we analyze the asymptotic behavior of the function $f\left(y,z\right)$, with $y \equiv \hbar \omega_{q,s}/k_B T$ and $z \equiv \hbar \gamma / k_B T $, to gain insight into how the temperature dependence of the resistivity evolves from the propagating to the diffusive regimes. Consider the extreme propagating regime,  where damping is absent, $\gamma = 0$. In this case, phasons behave as acoustic phonons and $f\left(y,z\right)$ becomes:
\begin{align}
\label{eq:f_propagating}
 f\left(y,z=0\right) = \frac{\pi}{y^2}+ \frac{\pi}{y^2} \left[ \frac{y^2}{4 \, \mathrm{sech}^2(y/2)} - 1 \right].
\end{align}
The first term corresponds to classical equipartition, and as such gives the standard linear-in-$T$ resistivity, $\rho \sim T$.\cite{DasSarma2008} It is dominant at temperatures that are high compared to $T_{\textrm{BG}}$, $y \ll 1$, in which case the second term vanishes. For $y \gg 1$, which corresponds to $T \ll T_{\textrm{BG}}$, one finds the well-known $\rho \sim T^4$ behavior (for a circular Fermi surface), as obtained for electron-acoustic phonon scattering in graphene.\cite{DasSarma2008,Mauri2014}

What happens once $\gamma$ increases and we move toward the diffusive regime? As long as $T_{\gamma} < T_{\textrm{BG}}$, the temperature scale where linear-in-$T$ resistivity emerges remains $T_{\textrm{BG}}$, since deviation from classical equipartition is driven by electrons being scattered off of propagating phason modes. However, a new linear-in-$T$ crossover temperature $T^{*}$ emerges when $T_{\gamma} > T_{\textrm{BG}}$, since in this case the scattering phason modes are essentially all overdamped at $T_{\textrm{BG}}$. In the asymptotic regime of $T_{\gamma} \gg \{ T, \,T_{\textrm{BG}} \}$, the function $f\left(y,z\right)$ becomes:

\begin{equation}
\label{eq:f_diffusive}
 f\left(y,z \gg \{ 1, y \}\right) \approx \frac{\pi}{y^2}+ \frac{2\pi}{y^2} \left[ \frac{1}{v^2} \, \psi_1 \left( 1 + \frac{1}{v} \right) - \frac{1}{v} \right]
\end{equation}   
where we defined the variable $v \equiv 2 \pi z/y^2$. This is the same expression one would have obtained for a purely diffusive response (i.e. dropping $\omega^2$ in the denominator of Eq. \ref{eq:susceptibility}). In contrast to Eq. \eqref{eq:f_propagating}, deviation from classical equipartition is now governed by the combined variable $v$, since the second term vanishes for $v \ll 1$. Therefore, the crossover temperature $T^{*}$ for the establishment of linear-in-$T$ resistivity (i.e. classical equipartition of the phason modes) can be estimated from the condition $T_\gamma T/ T_{\textrm{BG}}^2 \sim 1$, which gives
$T^{*} \sim \frac{T_{\textrm{BG}}^2}{T_{\gamma}}\ll T_{\textrm{BG}}$. 
The last inequality follows from the fact that, in the diffusive regime, $T_{\gamma} \gg T_{\textrm{BG}}$. Therefore, compared to the propagating regime, the temperature range across which $\rho \sim T$ extends to much lower temperatures, well below the Bloch-Gr\"uneisen temperature. This is the main result of our paper, which we confirm with an explicit calculation of the resistivity below.

It is not only the deviation from classical equipartition that is affected by the change in the character of the phason modes from propagating to overdamped. At the lowest temperatures, $T \ll T_{\gamma}, \, T_{\textrm{BG}}$, electron-phason scattering is always dominated by processes involving the low-energy part of the phason spectral weight, which in turn corresponds to the incoherent (i.e. overdamped) modes. Mathematically, it turns out that, regardless of the value of $T_{\gamma}/T_{\textrm{BG}}$, we can approximate $f(y\gg1,z)\approx 2\pi^2 z/3y^4$. As we show below, this gives rise to a $\rho \sim T^2$ behavior at the lowest temperatures. In the diffusive regime the temperature scale below which this behavior appears is the same $T^{*}$ obtained above. However, in the propagating regime, a new temperature scale $T^{**} \sim \sqrt{T_{\gamma} T_{\textrm{BG}}}$ emerges, with $T_{\gamma} \ll T^{**} \ll T_{\textrm{BG}}$, signaling the crossover from the characteristic acoustic-phonon driven behavior $\rho \sim T^4$ to the phason-driven behavior $\rho \sim T^2$.

\section{Explicit solution of the Boltzmann equation}

\label{sec:resistivity}

\subsection{Low-energy Dirac model: Relaxation-time approximation}
 
To proceed, we need the electron-phason coupling $g_s\left(\mathbf{k}_1,\mathbf{k}_2\right)$, which requires a model. 
For the electrons, we assume in this section a $\mathbf{k}\cdot\mathbf{p}$ description of the flat bands consisting of a Dirac Hamiltonian $\hat{\mathcal{H}}_{\textrm{e}}=v_F^*\,\boldsymbol{\hat{\Sigma}}\cdot(-i\hbar\boldsymbol{\partial})$ for each spin and valley defined around each corner of the hexagonal moir\'e Brillouin zone or \textit{moir\'e-valleys}, $\boldsymbol{\kappa}_{1,2}$. 
Omitting the spin, the Hamiltonian acts on a 8-component electronic wave function of the form $\boldsymbol{\psi}=(\vec{\psi}_{+,\boldsymbol{\kappa}_1},\vec{\psi}_{-,\boldsymbol{\kappa}_1},\vec{\psi}_{+,\boldsymbol{\kappa}_2},\vec{\psi}_{-,\boldsymbol{\kappa}_2})^T$, where $\vec{\psi}_{\zeta,\boldsymbol{\kappa}_i}=(\psi_{1,\zeta,\boldsymbol{\kappa}_i},\zeta\psi_{2,\zeta,\boldsymbol{\kappa}_i})^T$ are Dirac spinors of opposite chirality on each valley, $\zeta=\pm 1$, written in a basis of Bloch wave functions at points $\boldsymbol{\kappa}_i$ with complex eigenvalues under C$_{3z}$ rotations. Operators $\hat{\Sigma}_i$ are Pauli matrices acting on the Dirac spinors. Similarly, we can introduce Pauli matrices $\hat{\Gamma}_i$, $\hat{\Lambda}_i$ acting on valley and mini-valley degrees of freedom. These matrices provide a representation for the rest of operations in $D_6$,\cite{excitonic} the point group describing TBG. In this notation, the various symmetry-allowed electron-phason couplings are given by the Hamiltonian \cite{phasonsI} \begin{align}
& \hat{\mathcal{H}}_{\textrm{e-p}}= g_{A_1}\boldsymbol{\nabla}\cdot\boldsymbol{u}\,\hat{1}+g_{A_2}\left(\boldsymbol{\nabla}\times\boldsymbol{u}\right)_z\hat{\Lambda}_z\hat{\Gamma}_z \\
\nonumber
 & + g_{E_2}^{(1)}\left[\left(\partial_x u_y+\partial_y u_x\right)\hat{\Sigma}_x\hat{\Gamma}_z+ \left(\partial_xu_x-\partial_y u_y\right)\hat{\Sigma}_y\hat{\Gamma}_z\right]\\
 & + g_{E_2}^{(2)}\left[\left(\partial_x u_x-\partial_y u_y\right)\hat{\Sigma}_x\hat{\Lambda}_z- \left(\partial_xu_y+\partial_y u_x\right)\hat{\Sigma}_y\hat{\Lambda}_z\right],
\nonumber
\end{align}
where phason fluctuations are parametrized in terms of a collective coordinate $\boldsymbol{u}(\mathbf{r},t)$ describing long-wavelength transverse or longitudinal vibrations of the moir\'e pattern as a whole.\cite{phasonsI,phasons_Eslam}

The subscripts of the four coefficients $g_i$ refer to different irreducible representations of the $D_6$, and thus correspond to couplings to different lattice vibration patterns. While the contributions of each coupling to the resistivity can be summed up following Matthiessen's rule, symmetry dictates that they share the same temperature dependence. Therefore, hereafter we focus only on the $g_{A_2}$ term, which is expected to be the dominant one.\cite{phasonsI} Microscopically, this mode corresponds to a relative expansion/contraction of one layer with respect to the other, which is manifested as a transverse acoustic vibration of the moir\'e superlattice.


Considering only scattering within a single Fermi surface around each moir\'e-valley parametrized as $\mathbf{k}=k_F(\cos\theta,\sin\theta)$, and using $|g_{T}(\mathbf{k}_1,\mathbf{k}_2)|^2=g_{A_2}^2k_F^2\sin^2(\theta_1-\theta_2)$ deduced from this model, the resistivity within the relaxation-time approximation can be written as\begin{align} \label{eq:resistivity}
\rho=\rho_0\,I\left(t,\tau\right),\,\,\textrm{with}\,\,\rho_0=\frac{h}{e^2}\times\frac{g_{A_2}^2k_F^2}{4\varrho \left(v_F^*\right)^2 k_BT_{\textrm{BG}}},
\end{align}
and where we introduced the reduced temperature $t \equiv \frac{T}{T_{\textrm{BG}}}$ and the ratio $\tau \equiv \frac{T_{\gamma}}{T_{\textrm{BG}}}$. The dimensionless function $I(t,\tau)$ contains the remaining momentum integral in the inverse transport time, and is given by: \begin{align} \label{eq:I}
I\left(t,\tau\right)=\frac{16}{\pi^2 t}\int_0^1 du\,u^4\sqrt{1-u^2}\,f\left(\frac{u}{t},\frac{\tau}{t}\right).
\end{align}
Note that the integrand contains additional terms arising from the suppression of forward-scattering processes and the momentum dependence of the electron-phason coupling. 

\begin{figure}[t!]
\begin{center}
\includegraphics[width=\columnwidth]{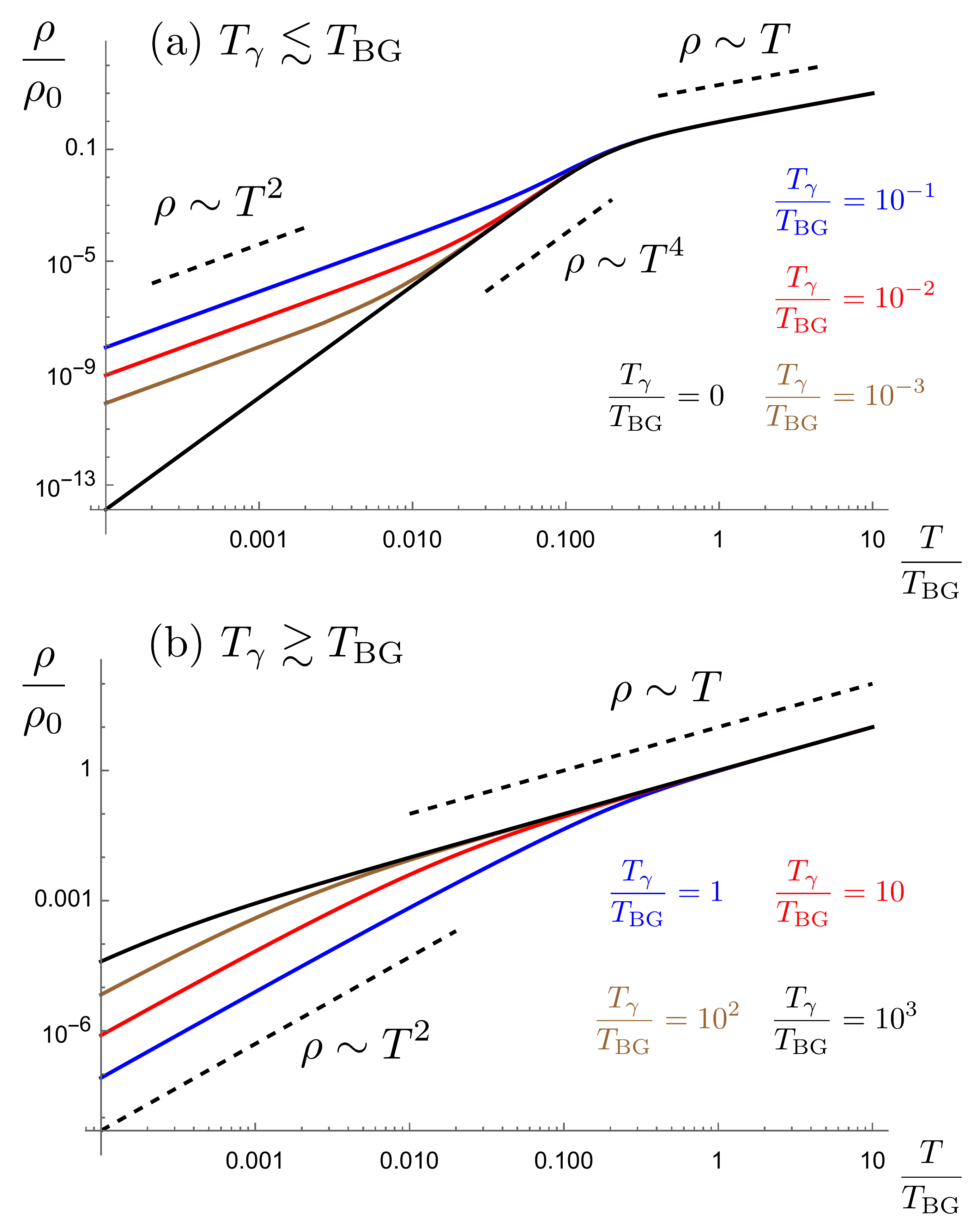}
\caption{Temperature dependence of the resistivity in the Dirac approximation for the flat bands obtained from a relaxation-time approximation solution of the Boltzmann equation. Numerical evaluation of $I(t,\tau)$ as a function of $t= T/T_{\textrm{BG}}$ for fixed values of $\tau = T_{\gamma}/T_{\textrm{BG}}$ in the propagating (panel a) and  diffusive (panel b) regimes. Both plots are in logarithmic scale. In the diffusive regime, the linear-in-$T$ resistivity extends down to a new (smaller) scale $T^{*}\ll T_{\textrm{BG}}$.} 
\label{fig:transport}
\end{center}
\end{figure}


Using the asymptotic expansions for $f(y,z)$ discussed in the previous section, it is straightforward to obtain the asymptotic temperature dependencies of the resistivity in different limits. In the propagating regime, $T_{\gamma} \ll T_{\textrm{BG}}$, we obtain
\begin{align}
\label{eq:resistivity_analitic}
\rho
\approx\rho_0\times
\begin{cases}
 \frac{T}{T_{\textrm{BG}}} & \textrm{if}\,\,T\gg T_{\textrm{BG}},\\
  \frac{64\pi^3}{15}\left(\frac{T}{T_{\textrm{BG}}}\right)^4 & \textrm{if}\,\,T^{**} \ll T\ll T_{\textrm{BG}},\\
\frac{8\pi}{3}\frac{T_{\gamma}}{T_{\textrm{BG}}}\left(\frac{T}{T_{\textrm{BG}}}\right)^2 & \textrm{if}\,\,T\ll T^{**}.
\end{cases}
\end{align}
The asymptotic behaviors above $T^{**}$ are the same as in the case of acoustic-phonon scattering,\cite{DasSarma,Adam1,Adam2} displaying a crossover from linear-in-$T$ resistivity to $\rho \sim T^4$ upon crossing $T_{\textrm{BG}}$. The low-temperature behavior $\rho \sim T^2$ arises from the contribution from the diffusive phason modes, which dominate at low $T$. The crossover temperature $T^{**}$ can be estimated by comparing the latter with the  Bloch-G\"uneisen contribution to the resistivity, yielding\begin{align}
T_{**}=\sqrt{\frac{5T_{\gamma}T_{\textrm{BG}}}{8\pi^2}}.
\end{align}

In the diffusive regime, $T_{\gamma}\gg T_{\textrm{BG}}$, we can use the asymptotic form for $f(y,z)$ in Eq.~\eqref{eq:f_diffusive}. We find $I(t,\tau)\approx t\mathcal{J}(2\pi t \tau)$ with \begin{align}
\mathcal{J}\left(\tilde{t}\right)=1-\frac{1}{\tilde{t}}+\frac{32}{\pi \tilde{t}^2}\int_0^1 dy\, y^6\sqrt{1-y^2}\,\psi_1\left(1+\frac{y^2}{\tilde{t}}\right).
\end{align}
As anticipated in Sec.~\ref{sec:transport}, the asymptotic behaviors of the resistivity are governed by a single-argument function. The argument can be interpreted as a new reduced temperature $\tilde{t}\equiv  2\pi t \tau =T/T^*$, with the new characteristic temperature scale in this regime given by
\begin{align} \label{eq:crossover}
T^{*}=\frac{T_{\textrm{BG}}^2}{2\pi T_{\gamma}}.
\end{align}
Using the results $\mathcal{J}(\tilde{t}\gg1)\approx 1$ and $\mathcal{J}(\tilde{t}\ll1)\approx 4 \tilde{t}/3$, we find the asymptotic behaviors of the resistivity \begin{align}
\rho\approx\rho_0\times\begin{cases}
\frac{T}{T_{\textrm{BG}}} & \textrm{if}\,\,T\gg T^{*},\\
\frac{4}{3}\frac{T^{2}}{T^{*}T_{\textrm{BG}}}
 & \textrm{if}\,\,T\ll T^{*}.
\end{cases}
\end{align}
Therefore, as anticipated, $T^* \ll T_{\textrm{BG}}$ is the new crossover temperature above which the resistivity is linear in $T$.

The schematic phase diagram in Fig. \ref{fig:fig1} is built based on the asymptotic behaviors derived here. To further confirm them, we numerically evaluated the function $I\left(t,\tau\right)$ in Eq. \eqref{eq:I}, which fully determines the temperature dependence of the resistivity in Eq. \eqref{eq:resistivity}. Figure~\ref{fig:transport}(a) shows $I\left(t,\tau\right)$ in the propagating regime, highlighting the crossover from linear-in-$T$ to $T^4$ at about $T_{\textrm{BG}}$, followed by another crossover to $T^2$ at temperatures between $T_\gamma$ and $T_{\textrm{BG}}$. In the diffusive regime, shown in Fig. \ref{fig:transport}(b), the linear-in-$T$ behavior extends to temperatures well below $T_{\textrm{BG}}$ for large enough $T_\gamma$, confirming the main result of our analysis. Moreover, as shown in this figure, collisions with phasons give rise to a large resistivity at very low temperatures, no longer limited by the Bloch-Gr\"uneisen temperature.  

\subsection{Tight-binding model: Beyond the relaxation-time approximation}

In order to verify that the extended linear-in-$T$ resistivity is not an artifact of the relaxation-time approximation or the low-energy Dirac approximation, we also computed numerically the resistivity for different electron fillings in a six-band tight-binding model of the bands of TBG at the magic angle.\cite{Po2019} The total Hamiltonian for one valley can be written as $\hat{H}=\hat{H}_0+\hat{H}_{\textrm{e-ph}}$, where the first term is the band Hamiltonian, $\hat{H}_0=\sum_{\mathbf{k}}\hat{\boldsymbol{\Psi}}_{\mathbf{k}}^{\dagger}\hat{\mathcal{H}}_{\mathbf{k}}\hat{\boldsymbol{\Psi}}_{\mathbf{k}}$, written in the following basis of fermion operators:\begin{align}
\hat{\boldsymbol{\Psi}}_{\mathbf{k}}=\left(\hat{p}^z_{\mathbf{k}},\hat{p}^+_{\mathbf{k}},\hat{p}^-_{\mathbf{k}},\hat{s}^1_{\mathbf{k}},\hat{s}^2_{\mathbf{k}},\hat{s}^3_{\mathbf{k}}\right)^T.
\end{align} The operators in the first three entries correspond to orbitals with $p^z$ and $p^{\pm}=p^x\pm ip^y $ symmetry defined on the triangular lattice formed by the moir\'e beating pattern maxima (regions of local AA stacking). The other three operators correspond to orbitals with $s$ symmetry defined on the Kagome lattice formed by the points half-way between the maxima. The form of the matrix Hamiltonian $\hat{\mathcal{H}}_{\mathbf{k}}$ in this basis (including the values of the tight-binding parameters employed in the calculation) can be found in Ref.~\onlinecite{Po2019}. The second term in the Hamiltonian is the electron-phason coupling, which is modelled for simplicity as a deformation potential of strength $g$ diagonal in orbital indices:\begin{align}
    \hat{H}_{\textrm{e-ph}}=\frac{i g}{\sqrt{A}}\sum_{\mathbf{k},\mathbf{q}}\left|\mathbf{q}\right|u_{L}\left(\mathbf{q}\right)\hat{\boldsymbol{\Psi}}^{\dagger}_{\mathbf{k}+\mathbf{q}}\hat{\boldsymbol{\Psi}}_{\mathbf{k}},
\end{align}where $A$ is the total area of the system. We only consider here the coupling with longitudinal phasons.

For the resistivity calculation, to go beyond the relaxation-time approximation, we employed a variational approach. The idea is to expand the solution of Eq.~\eqref{eq:Boltzmann_eq} in a basis of trial functions $\Phi_{\mathbf{k}}^{(m)}$, $\Phi_{\mathbf{k}}=\sum_{m}\eta_{m}\Phi_{\mathbf{k}}^{(m)}$. According to the variational principle,\cite{Ziman} the coefficients $\eta_{m}$ can be determined by minimization of the resistivity in Eq.~\eqref{eq:resistivity1} understood now as a functional defined on the space of trial functions.

We focus on band fillings $\nu$ (defined as the number of electrons per moir\'e supercell) such that the Fermi surface consists on a single contour centered at the $\Gamma$ point of the moir\'e Brillouin zone. The trigonal distortion of the Fermi contours compels us to look for variational solutions beyond the relaxation-time approximation, $\Phi_{\mathbf{k}}^{(1)}\propto\hat{\mathbf{u}}_{\mathbf{E}}\cdot\mathbf{k}$. This is only an exact solution for isotropic Fermi surfaces, which is not the case beyond the Dirac approximation. Hereafter $\mathbf{k}$ corresponds to a point in the Fermi surface parametrized as $\mathbf{k}=k_F(\theta)(\cos\theta,\sin\theta)$, where $\theta$ is the polar angle measured with respect to $\hat{\mathbf{u}}_{\mathbf{E}}=\hat{\mathbf{x}}$ and $k_F(\theta)$ is obtained directly from the six-band tight-binding model. A generalization of the relaxation-time ansatz consists of an expansion in angular harmonics of the form $\cos(m\theta)$, $\sin(m\theta)$, with $m$ an integer. We can use the symmetries of TBG to restrict this expansion. The model possesses full $D_3$ point group symmetry within a single valley. In particular, C$_{2x}$ symmetry exchanging layers ($\theta\rightarrow-\theta$) forbids $\sin(m\theta)$ terms, since the associated variational integrals cancel on the Fermi surface. Moreover, C$_{3z}$ symmetry implies that harmonics $\cos(m\theta)$ with $m=0_{\textrm{mod}3}$ do not contribute either. Thus, we restrict the variational solution to a set of trial functions of the form $\Phi_{\mathbf{k}}^{(m)}\propto\cos(m\theta)$ with $m=1,2,4,5...\, p$. 

In a variational calculation with an expansion in angular harmonics up to order $p$, the minimum of the resistivity in Eq.~\eqref{eq:resistivity1} can be written as 
\begin{align}
\label{eq:resistivity2}
\rho=\rho_0\,I_{p}\left( t,\tau\right),\,\,\textrm{with}\,\,\rho_0=\frac{\hbar}{e^2}\times\frac{g^2k_{\textrm{BG}}^2}{4\varrho \left\langle \boldsymbol{v}_{k_F}^2 \right\rangle k_BT_{\textrm{BG}}}.
\end{align}
Note that the prefactor $\rho_0$ now contains $k_{\textrm{BG}}$, which corresponds to the maximum momentum exchanged between electrons on the Fermi surface, $k_BT_{\textrm{BG}}=\hbar c_L k_{\textrm{BG}}$, defining the Bloch-Gr\"uneisen scale for scattering with longitudinal phasons. Moreover, $\rho_0$ also contains $\langle \boldsymbol{v}_{k_F}^2 \rangle$, which is the average of the squared group velocity on the Fermi surface,\begin{align}
\left\langle \boldsymbol{v}_{k_F}^2 \right\rangle\equiv \int_0^{2\pi} d\theta\,\left|\boldsymbol{v}_{k_F(\theta)}\right|^2.
\end{align}
Finally, $I_p(t,\tau)$ is a dimensionless function containing the temperature dependence of the resistivity,\begin{align}
I_p\left(t,\tau\right)\equiv\frac{1}{\vec{X}^T\cdot\hat{P}^{-1}\left(t,\tau\right)\cdot\vec{X}}.
\end{align}
Here, $\vec{X}=(X_1,X_2... X_p)^T$ are vectors whose components correspond to variational integrals on the Fermi surface of the form\begin{align}
 X_m=\int_0^{2\pi} d\theta\,\frac{k_F\left(\theta\right)\boldsymbol{v}_{k_F(\theta)}^x}{\left|\boldsymbol{v}_{k_F(\theta)}\right|}\,\cos\left(m\theta\right).
\end{align}
The components of the matrix $\hat{P}(t,\tau)$ in this basis for each value of the reduced temperature $t$ and parameter $\tau$ read \begin{widetext}\begin{align}
P_{nm}=\left\langle \boldsymbol{v}_{k_F}^2 \right\rangle\int d\theta_1\int d\theta_2\,\frac{k_F\left(\theta_1\right)k_F\left(\theta_2\right)\left|\mathbf{k}_1-\mathbf{k}_2\right|^2\left[\cos\left(n\theta_1\right)-\cos\left(n\theta_2\right)\right]\left[\cos\left(m\theta_1\right)-\cos\left(m\theta_2\right)\right]}{x k_{\textrm{BG}}^2\left|\boldsymbol{v}_{k_F(\theta_1)}\right|\left|\boldsymbol{v}_{k_F(\theta_2)}\right|}\,f\left(\frac{\left|\mathbf{k}_1-\mathbf{k}_2\right|}{t\,k_{\textrm{BG}}},\frac{\tau}{t}\right),
\end{align}
\end{widetext} 
where $f(y,z)$ is the same function defined in Eq.~\eqref{eq:f}. For fixed values of $p$, electronic filling $\nu$, and temperature arguments $t$ and $\tau$, we computed the integrals on the Fermi surface numerically and then inverted the matrix $\hat{P}$ to obtain the resistivity.

\begin{figure}[t!]
\begin{center}
\includegraphics[width=\columnwidth]{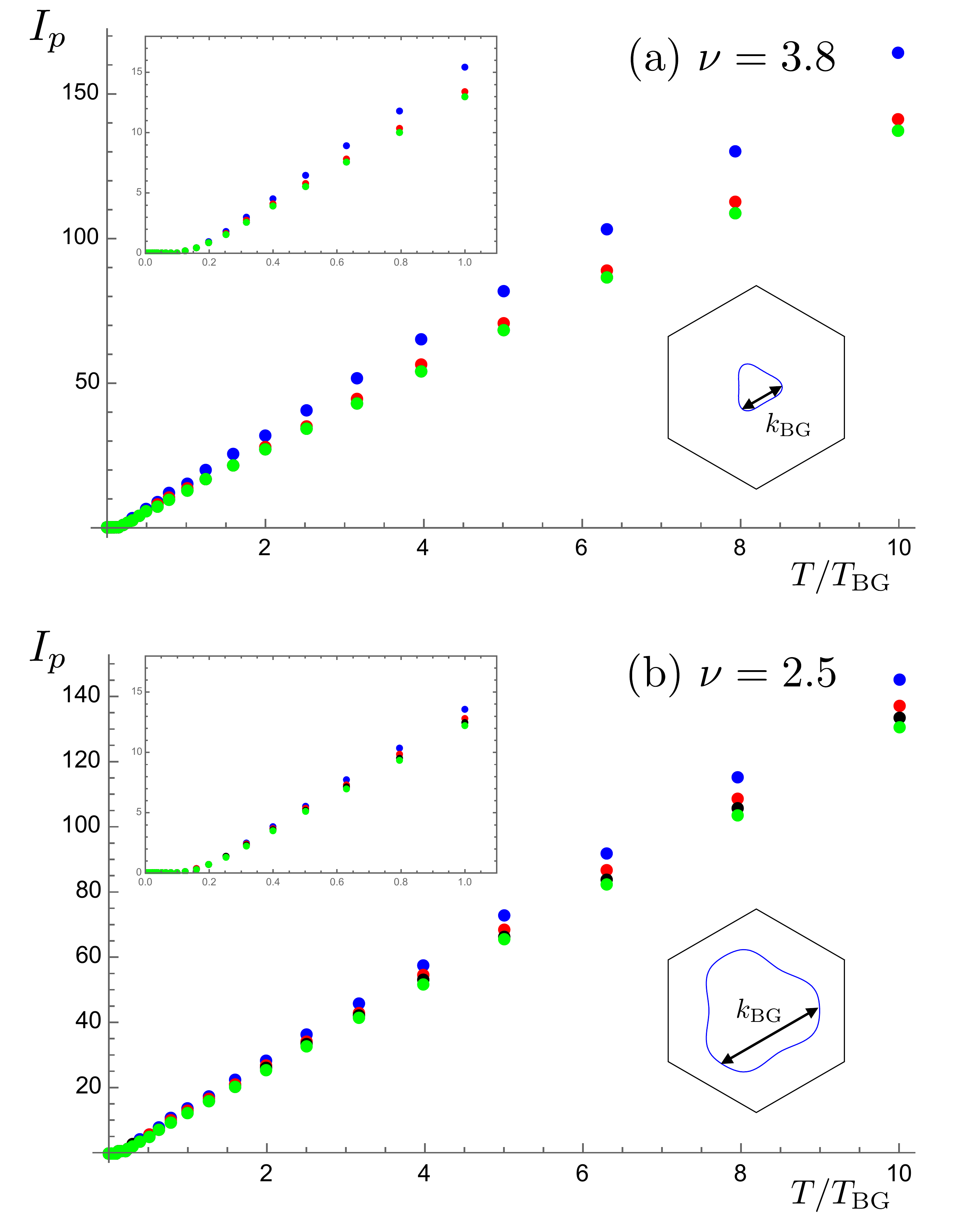}
\caption{Numerical evaluation of $I_p(t,\tau)$ in Eq. (\ref{eq:resistivity2}) as a function of the first argument, i.e. temperature $T$ in units of $T_{\textrm{BG}}$). Panel (a) shows the results for an electron concentration of $n=2.21\times 10^{12}$ cm$^{-2}$, and panel (b) for $n=1.45\times 10^{12}$ (both at the magic angle). Blue points correspond to the scattering-time approximation, $p=1$, red points to $p=2$, black points to $p=4$, and green points to $p=5$. The second argument is fixed to $\tau=T_{\gamma}/T_{\textrm{BG}}=10^{-2}$ in all cases. The insets in the upper side of the plots represent the values of the resistivity for temperatures lower than $T_{\textrm{BG}}$. The insets in the lower side of the plots represents the Fermi surfaces in one of the valleys for that particular filling as obtained from the six-band tight-binding model of Ref. \onlinecite{Po2019}.} 
\label{fig:numerics}
\end{center}
\end{figure}

\begin{figure}[t!]
\begin{center}
\includegraphics[width=\columnwidth]{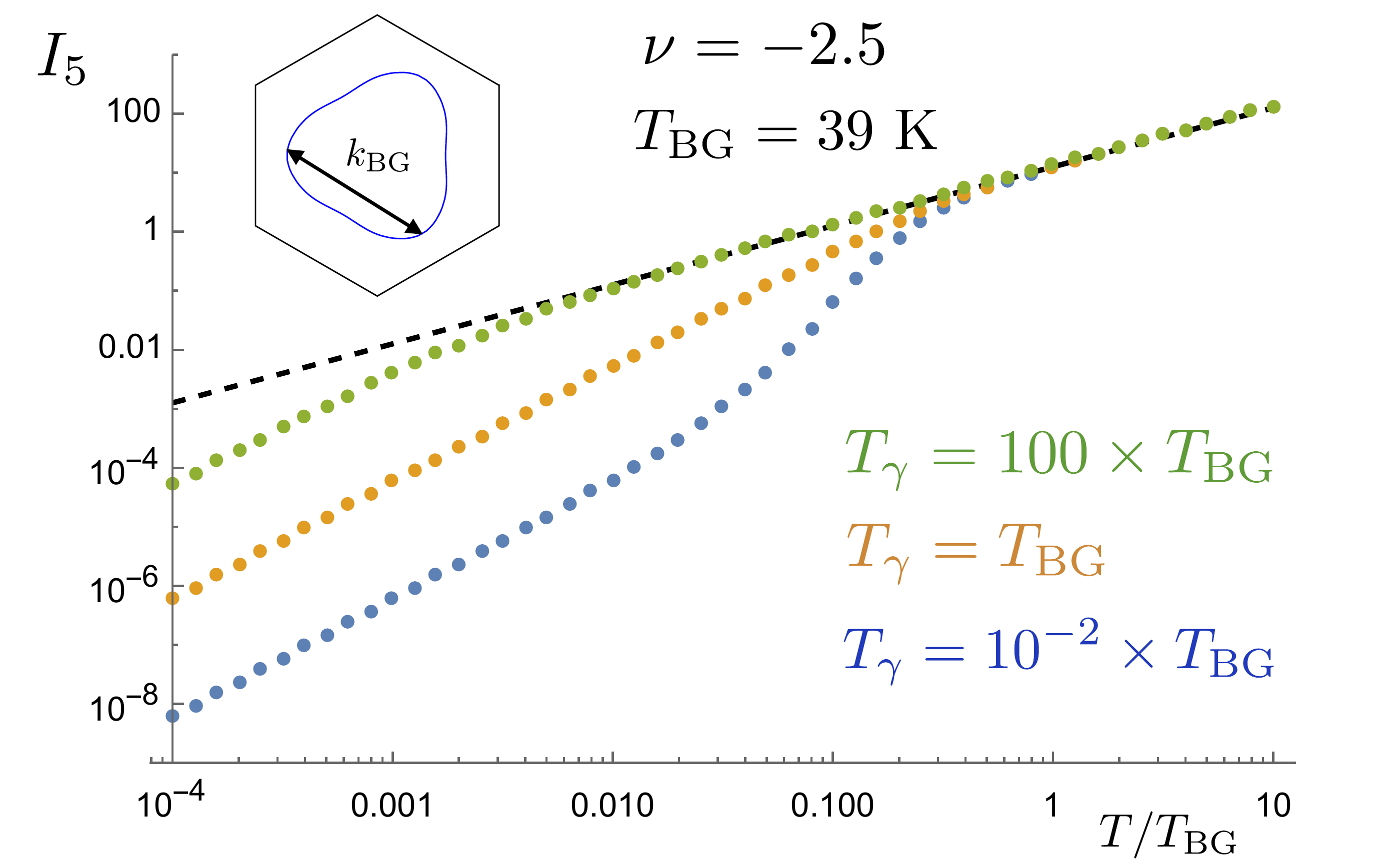}
\caption{Numerical calculation of the resistivity in Eq. (\ref{eq:resistivity2}) as a function of reduced temperature $T / T_{\textrm{BG}}$. A variational trial function up to the $p=5$ harmonic was considered. The blue points correspond to the resistivity in the propagating regime ($T_{\gamma}=10^{-2}\times T_{\textrm{BG}}$), the green points to the diffusive regime ($T_{\gamma}=10^{2}\times T_{\textrm{BG}}$), and the orange points to an intermediate situation ($T_{\gamma}=T_{\textrm{BG}}$). For reference, the black dashed line corresponds to the equipartition limit. The hole concentration is $n=1.45\times 10^{12}$ cm$^{-2}$ in all cases.  Note that both axes are in logarithmic scale. The inset shows the Fermi surface and the magnitude of $k_{\textrm{BG}}$ for this filling as obtained from the six-band tight-binding model of Ref. \onlinecite{Po2019}.} 
\label{fig:resistivity}
\end{center}
\end{figure}

Figure~\ref{fig:numerics} shows the numerical calculation of $I_p(t,\tau)$ as a function of the reduced temperature for two representative fillings of the conduction band, corresponding to electron densities of $n=2.21\times 10^{12}$ cm$^{-2}$ [panel (a), filling $\nu=3.8$] and  $n=1.45\times 10^{12}$ cm$^{-2}$ [panel (b), filling $\nu=2.5$]; recall that $\nu=0$ corresponds to charge-neutrality and $\nu=4$ to a fully filled ``flat'' band. 
The insets show the corresponding Fermi contours in one of the valleys as well as the magnitude of $k_{\textrm{BG}}$. In all cases, we set $T_{\gamma}/T_{\textrm{BG}}=10^{-2}$. Taking $c_L=1.5\times 10^4$ m/s, the corresponding Bloch-Gr\"uneisen scales for each filling are $T_{\textrm{BG}}=15$ K ($\nu=3.8$) and $T_{\textrm{BG}}=38$ K ($\nu=2.5$). The temperature dependence of $I_p(t,\tau)$ is that expected for propagating phasons: The resistivity is linear in $T$ down to $T_{\textrm{BG}}$, below which (see upper insets) it decreases quickly with a higher exponent. The different colours correspond to variational ansatzes with different number of harmonics. As $p$ increases the numerical values of the resistivity decreases, indicating that the variational solution improves. For the range of fillings and temperatures considered here, we find that the calculation with $p=5$ already provides good convergence. For the largest fillings, for which the shape of the Fermi surface is smoother, the convergence is actually quicker. Indeed, in Fig.~\ref{fig:numerics}(a), the green points, corresponding to $p=5$, overlap with the black points, corresponding to $p=4$.

Figure~\ref{fig:resistivity} shows the temperature dependence in double logarithmic scale for a filling $\nu=-2.5$, corresponding to a hole concentration of $n=1.45\times 10^{12}$ cm$^{-2}$ and a Bloch-Gr\"uneisen scale $T_{\textrm{BG}}=39$ K. Each color represents a different value of the damping parameter $\gamma$. The results are consistent with our discussion in Sec.~\ref{sec:transport}. In the propagating limit (blue points) the Bloch-Gr\"uneisen regime is replaced by $\rho\propto T^2$ at the lowest temperatures. As phason damping increases, the Bloch-Gr\"uneisen regime is washed out (orange points, $T_{\gamma}=T_{\textrm{D}}$). In the diffusive regime (green points) there is a direct crossover from the $\rho\propto T^2$ to the classical equipartition regime $\rho\propto T$ at $T^{*}\sim 10^{-2} T_{\textrm{BG}}$. These temperature regimes are consistently reproduced for the different fillings considered in our calculations. Moreover, numerical changes in the value of $I_5(t,\tau)$ for the same values of the arguments but different fillings are negligible in the logarithmic scale of Fig.~\ref{fig:resistivity}. This suggests that in the limit of classical equipartition the dependence on carrier concentration is dominated by the averaged Fermi velocity, which decreases with increasing filling away from half-filling in our calculations.


\section{Phason contribution to the specific heat}

\label{sec:heat_capacity}


Besides promoting electronic scattering processes, the transfer of phason spectral weight to lower energies has immediate consequences for thermodynamic  quantities as well. To illustrate this effect, we focus here on the phason contribution to the specific heat. In particular, we consider the specific heat at constant (hetero-)stress, \begin{align}
C_{\sigma}\equiv T\left(\frac{\partial S}{\partial T}\right)_{\sigma_{ij}},
\end{align}
which measures how entropy $S$ (here defined per moir\'e supercell) changes with temperature for a fixed value of the forces between layers. The contribution arising from a phonon mode $s$ is directly related to its density of states,
\begin{align}
C_{\sigma}=k_B\int_0^{\infty} d\omega\,\left(\frac{\hbar\omega}{2k_B T}\right)^2\sinh^{-2}\left(\frac{\hbar\omega}{2k_BT}\right)\mathcal{D}_{s}\left(\omega\right).
\end{align}
Here, $\mathcal{D}_{s}\left(\omega\right)$ is defined as\begin{align}
\mathcal{D}_{s}\left(\omega\right)=\frac{2\varrho A_{\textrm{m}}}{\pi \omega}\int \frac{d\mathbf{q}}{(2\pi)^2}\,\omega_{s,\mathbf{q}}^2\,\chi_s''\left(\mathbf{q},\omega\right),
\end{align}
where $A_{\textrm{m}}$ is the area of the moir\'e supercell and the integration is over the moir\'e Brillouin zone.

In our low-energy description, we cut-off the linear dispersion relation $\omega_{s,\mathbf{q}}=c_s|\mathbf{q}|$ at the Debye momentum $q_{\textrm{D}}=2\sqrt{\pi/A_{\textrm{m}}}$. The density of states (per moir\'e supercell) of mode $s$ reads then\begin{align}
 \mathcal{D}_s\left(\omega\right)=& \frac{2}{\omega_{\textrm{D}}}\,g\left(\frac{\omega}{\omega_{\textrm{D}}},\frac{\gamma}{\omega_{\textrm{D}}}\right),
\end{align}
where $\omega_{\textrm{D}}=c_sq_{\textrm{D}}$ is the associated Debye frequency and
\begin{align}
 & g\left(x,y\right)\equiv\frac{y}{\pi}\int_0^1dz\frac{z}{\left(z-x^2\right)^2+x^2y^2}=\\
 & \frac{y}{2\pi}\,\ln\frac{\left(x^2-1\right)^2+x^2y^2}{x^4+x^2y^2}+\frac{x}{\pi}\left[\textrm{arccot}\frac{y}{x}+\arctan\frac{1-x^2}{xy}\right].
\nonumber
\end{align}
This function interpolates between the expected linear-in-energy density of states for phonons in the total absence of damping, $g(x,0)=x$, and the asymptotic limit in which all modes are overdamped, \begin{align}
\label{eq:g_diff}
g\left(x,y\gg  \{ 1, x \} \right)\approx\frac{y}{2\pi}\ln\left(1+\frac{1}{x^2y^2}\right).
\end{align}
For intermediate damping values, the density of states is non-monotonic. The specific heat can be written as\begin{subequations}\begin{align}
& C_{\sigma}=k_B\,h\left(t,\tau\right),\,\,\textrm{with}\\
& h\left(t,\tau\right)\equiv\frac{t}{2}\int_0^{\infty} dz\,\frac{z^2}{\sinh^2\frac{z}{2}}\,g\left(tz,\tau\right).
\end{align}
\end{subequations}
The reduced temperature $t=T/T_{\textrm{D}}$ is now defined with respect to the Debye temperature, $T_{\textrm{D}}=\hbar\omega_{\textrm{D}}/k_B$, and similarly $\tau=T_{\gamma}/T_{\textrm{D}}$.

\begin{figure}[t!]
\begin{center}
\includegraphics[width=\columnwidth]{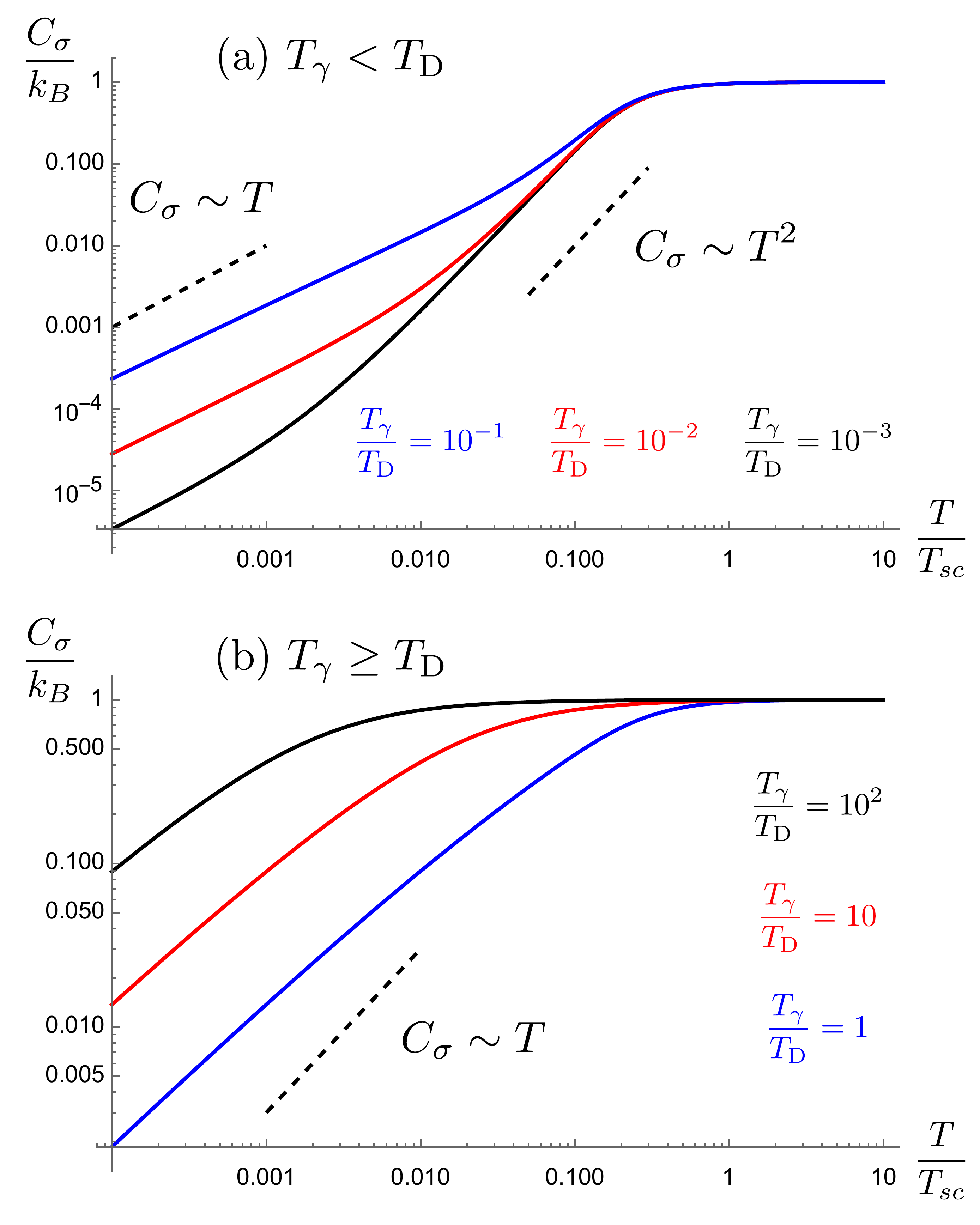}
\caption{Temperature dependence of the specific heat. Numerical evaluation of $C_\sigma / k_B = h(t,\tau)$ as a function of the reduced temperature $t=T/T_{\textrm{D}}$ for different values of $\tau=T_{\gamma}/T_{\textrm{D}}$ in the propagating [panel (a), $T_{\gamma}<T_{\textrm{D}}$] and  diffusive [panel (b), $T_{\gamma}>T_{\textrm{D}}$] regimes. Both plots are in logarithmic scale.} 
\label{fig:cp}
\end{center}
\end{figure}

Using the limiting behaviors of $g(x,y)$ above, we obtain the asymptotic temperature dependencies of the specific heat. In the propagating regime, $T_{\gamma} \ll T_{\textrm{D}}$, we obtain\begin{align}
C_{\sigma}\approx k_B\times\begin{cases}
1 & \textrm{if}\,\, T\gg T_{\textrm{D}},\\
12\,\zeta\left(3\right)\left(\frac{T}{T_{\textrm{D}}}\right)^2 & \textrm{if}\,\, T_{\textrm{D}}\gg T\gg T_{\gamma},\\
\frac{2\pi T_{\gamma}}{3T_{\textrm{D}}}\frac{T}{T_{\textrm{D}}}\,\ln\left(\frac{T_{\textrm{D}}^2}{T_{\gamma} T}\right) & \textrm{if}\,\, T\ll T_{\gamma},
\end{cases}
\end{align}
where $\zeta(x)$ is the Riemann zeta function. In addition to the expected crossover from Dulong-Petit to Debye behaviors at temperatures of the order of $T_{\textrm{D}}$, there is a new crossover to $C_{\sigma}\propto T$ dominated by incoherent phasons below the scale $T_{\gamma}$. This is an interesting result, since a linear-in-$T$ specific heat is characteristic of a metal. Here, however, it is a consequence of the overdamped nature of the phasons at low energies, and would emerge even in the correlated insulating phase of TBG. We note that the logarithmic pre-factor comes from the divergence of the phason density of states at $\omega=0$, which could be regularized by an infrared cut-off for the theory, as given, for example, by the disorder pinning length.\cite{phasonsII}

In the diffusive regime ($T_{\gamma} \gg T_{\textrm{D}}$), the intermediate phonon-like $T^2$ regime is washed out. There is a single crossover from classical equipartition to the incoherent regime, where $C_{\sigma}\propto T$, corresponding to a smaller temperature scale $T_{\textrm{D}}^2/T_{\gamma}<T_{\textrm{D}}$. Similarly to our calculation for the resistivity above, this crossover scale can be identified by introducing a new reduced temperature variable $\tilde{t}\equiv t \tau$ in the expression for $C_{\sigma}$ as given by the diffusive limit of the density of states, Eq.~\eqref{eq:g_diff}. 

To verify these analytical expressions we numerically evaluated the function $h(t,\tau)$. Figure~\ref{fig:cp} shows $h(t,\tau)$ as a function of the reduced temperature for different values of damping. In the propagating regime [panel (a)], we distinguish the three different regimes obtained analytically. In the diffusive regime [panel (b)], the crossover from the classical to the equipartition regime is direct and there is no phonon-like $T^2$ dependence.


Measuring the specific heat of TBG would be extremely challenging, which makes a direct verification of these predictions difficult. Nevertheless, the fingerprints of overdamped phasons should also appear in other thermodynamic and transport quantities that are sensitive to the presence of low-energy bosonic modes. In this regard, thermal conductivity is an appealing observable, particularly in the correlated insulating phase of TBG, where low-energy mechanical excitations should give the leading contribution. While a rigorous calculation of thermal conductivity is beyond the scope of the present work, we provide here an estimate for the case in which lattice conduction is limited by disorder (impurities, the boundaries of the sample, etc.). In a relaxation-time approximation for the distribution function of phasons, the contribution of mode $s$ to the thermal conductivity can be written as\cite{Ziman}\begin{align}
\kappa=\frac{k_B c_s^2}{2A_{\textrm{m}}}\int_0^{\infty}d\omega\,\tau(\omega)\left(\frac{\hbar\omega}{2k_B T}\right)^2\sinh^{-2}\left(\frac{\hbar\omega}{2k_BT}\right)\mathcal{D}_{s}\left(\omega\right).
\end{align}
For elastic scattering we can assume a frequency-independent relaxation time, $\tau(\omega)\equiv \tau$. We arrive then at the usual kinetic formula for the thermal conductivity,\begin{align}
\kappa=\frac{\tau c_s^2C_{\sigma}}{2A_{\textrm{m}}},
\end{align}
where $C_\sigma$ is the phason contribution to the specific heat and $\tau c_s^2/2$ can be interpreted as the phason thermal diffusivity. 
Identifying $\tau^{-1}\sim \gamma$ and using the asymptotic expression for the specific heat at low temperatures, we arrive at the following expression for the thermal conductivity:\begin{align}
\kappa(T\ll T_{\gamma}) \approx\frac{k_B^2 T}{12\hbar}\,\ln\left(\frac{T_{\textrm{D}}^2}{T_{\gamma}T}\right),
\end{align}
which reveals a linear-in-temperature behavior.

\section{Discussion}

\label{sec:conclusions}

In this paper, we showed that electron-phason scattering can lead to a linear-in-$T$ resistivity down to temperatures much lower than the Bloch-Gr\"uneisen temperature. In this scattering mechanism, the momentum that the electrons yield to the moir\'e superlattice via collisions with its long-wavelength phason fluctuations is rapidly degraded through friction between the layers. The latter is a generic feature of incommensurate lattices,\cite{Winger,Rice,Rice_bis,Currat_etal,Cano_etal} 
parametrized here by the damping coefficient $\gamma$. 
Any form of dissipative coupling between the two layers contributes to $\gamma$, including stick-slip processes caused by disorder in the stacking arrangement. The existence of various possible mechanisms for damping makes it difficult to estimate $T_{\gamma}$. If the origin of $\gamma$ is mechanical, the natural scale is the one defined by the van der Waals forces between the layers. These are weak but non-negligible close to the magic angle, where the effects of lattice relaxation are substantial. The adhesion energy between carbon layers per unit area is of the order of 4 meV/$\AA^2$ according to \textit{ab initio} calculations.\cite{Carr2018} Integrated over graphene's unit cell, this gives $T_{\gamma} \approx 250$ K. This is an upper-bound estimate for $T_{\gamma}$, as the exact value should depend on the amount of disorder and tensions at the edges of the device. For a lower-bound estimate, we can take the typical values of the damping coefficient employed in molecular-dynamics simulations of tribological properties of defect-free graphene interfaces, which are of several $1/$ps.\cite{Wang_2023} This translates to a scale of $T_{\gamma}\sim 10$ K, of the same order as $T_{\textrm{BG}}$, placing TBG in the right-hand side of the diagram in Fig.~\ref{fig:fig1}. The key point is that inter-layer friction further extends to lower temperatures the regime of linear-in-$T$ resistivity down to $T^{*}\sim T_{\textrm{BG}}^2/T_{\gamma}$, which can be as small as $T^{*}\approx0.06$ K if we use the upper-bound estimate for $T_\gamma$.


Besides the resistivity, phason modes should also impact other transport properties, such as the thermal conductivity, as well as thermodynamic properties such as the specific heat. In particular, the phason contribution to the specific heat as function of temperature mimics the behavior of the resistivity. There are three different scaling behaviors with $T$ in the propagating regime and a single crossover from classical equipartion to an incoherent regime in the limit of diffusive phasons. Importantly, $C_{\sigma}\propto T$ at the lowest temperatures, $T\ll\textrm{min}\{T_{\gamma},T_{\textrm{D}}^2/T_{\gamma}\}$, which dominates over the contribution from acoustic phonons ($\propto T^2$). For the same reasons, we expect the thermal conductivity $\kappa$ to be dominated by phasons and change linearly with temperature under the appropriate conditions. Although the low-temperature behavior of $C_\sigma$ is that typical of metals, it arises from contributions from the mechanical, rather than the electronic degrees of freedom. As a result, a linear-in-$T$ specific heat is expected in TBG even in the insulating correlated phase.

Our results provide a solid framework for future studies to quantitatively assess the relevance of the electron-phason mechanism in addressing the puzzling linear-in-$T$ resistivity of TBG. Interestingly, this mechanism contains features of two scenarios invoked to explain this effect: electron-phonon scattering and quantum criticality. Of course, the linear-in-$T$ resistivity behavior discussed here is due to classical equipartition, rather than scattering by quantum critical fluctuations. However, at low temperatures, where electron-phason scattering leads to a $\rho \sim T^2$ behavior, the overdamped phasons are described by Eq. \eqref{eq:susceptibility}, and thus behave similarly to overdamped bosonic excitations typical of a metallic QCP.\cite{Millis1993,Abanov2003,Sachdev2010,Fradkin2001,Lohneysen_etal} In fact, the scattering function in Eq. \eqref{eq:f} is identical to that obtained for a metallic nematic QCP,\cite{Carvalho2019} except for the momentum dependence of the damping coefficient. This is because, in a quantum critical system, dissipation is due to electronic Landau damping, whereas here it is a purely mechanical effect. Moreover, while bosonic excitations are only gapless at the QCP, the phason spectrum is gapless everywhere -- although a small disorder-induced gap may emerge.\cite{phasonsII} This suggests that a low-temperature $\rho \sim T^2$ behavior may be more common in moir\'e superlattices. Interestingly, Ref.~\onlinecite{transport3} reported a quadratic-in-$T$ resistivity over certain doping ranges. 

The phason-based mechanism proposed here should be operative in other quasiperiodic structures. One example is the bismuth phase Bi-III at high pressure, which becomes a superconductor below $T_c\approx 7$ K and also displays linear-in-$T$ resistivity at low temperatures.\cite{Brown2018} Conversely, since this mechanism is specific to incommensurate lattices, it should be absent in graphene-based systems without a moir\'e superlattice. Recently, phenomena first observed in moir\'e systems, such as superconductivity and flavor-polarized metals, have also been reported in rhombohedral ABC graphene \cite{Zhou2021} and Bernal bilayer graphene.\cite{Zhou2022,Barrera2022} Since phasons are not present in these systems, it will be interesting to determine whether they display linear-in-$T$ resistivity down to temperatures lower that the Bloch-Gr\"uneisen scale.



\begin{acknowledgments}
H.O. acknowledges funding from the Spanish MCI/AEI/FEDER through Grant No. PID2021-128760NB-I00. R.M.F. was supported by the U.S. Department of Energy, Office of Science, Basic Energy Sciences, Materials Science and Engineering Division, under Award No. DE-SC0020045.
\end{acknowledgments}

\bibliography{references}

\end{document}